\documentclass[conference]{IEEEtran}
\IEEEoverridecommandlockouts
\usepackage{amssymb}
\usepackage{booktabs}

\usepackage{graphicx}

\newcommand{\BB}[1]{$[\![#1]\!]$}
\newcommand{\gro}{\mathcal{G}}

\begin{document}

\title{RingBP: Folding Code Automorphisms into Time for a Compact Real-Time Relay-BP Decoder}

\author{
\IEEEauthorblockN{
Junichiro Kadomoto
}
\IEEEauthorblockA{
\textit{The University of Tokyo}\\
Tokyo, Japan\\
kadomoto@mtl.t.u-tokyo.ac.jp
}
\thanks{This work has been submitted to the IEEE for possible publication. Copyright may be transferred without notice, after which this version may no longer be accessible.}
}

\maketitle

\begin{abstract}
\boldmath
Bivariate bicycle (BB) quantum LDPC codes realize a logical qubit with about ten times fewer physical qubits than surface codes, but they must be decoded in real time on a circuit-level decoding matrix with tens of thousands of edges. The prior real-time decoder runs Relay-BP, a belief-propagation algorithm, on this matrix fully in parallel, so its hardware cost grows with every edge of the matrix. We present RingBP, an architecture for partially parallel Relay-BP decoder ASICs. Our key observation is that the decoding matrix inherits the automorphism group of the code. For the gross code, its 30k-edge Tanner graph is a free $\mathbb{Z}_6\times\mathbb{Z}_{12}$ lift of a 426-edge base graph. RingBP maps the base graph onto space and sweeps the group in time over rotating check-node register tori, so every lift permutation becomes a fixed wire and a folding factor $P$ trades area for latency. The lift is found automatically from a circuit description in the Stim framework. In logic synthesis for a commercial 22\,nm CMOS technology, the gross-code decoder occupies 0.36--1.02\,mm$^2$ for $P$=1--12, which is 9--24\% of the fully parallel design, and meets a 1.05\,ns clock for every $P$. At $P=12$, one iteration takes 7.4\,ns at 0.50\,W, and a syndrome stream is decoded under a 32\,\textmu s deadline with no loss in logical error rate. For physical error rates up to 0.4\%, the logical error rate is on par with or below that of floating-point Relay-BP on three BB codes. These results show that turning the symmetry of the code into wiring yields compact, low-power real-time decoders for BB codes.
\end{abstract}

\IEEEpeerreviewmaketitle

\section{Introduction}
Large-scale quantum computers are believed to solve problems that classical computers cannot solve efficiently. Building such a computer requires quantum error correction, which encodes many physical qubits into a logical qubit and thereby suppresses its error rate~\cite{DaleyNature2022}. The code detects errors by repeatedly measuring parity checks on the physical qubits, and a classical decoder infers the errors from the measurement outcomes, called syndromes.

The syndromes of the error-correcting code must be decoded at high speed. If the decoder falls behind the syndrome stream, the backlog grows without bound~\cite{terhal2015}. For surface codes on superconducting qubits, many dedicated hardware decoders based on matching~\cite{liyanage2023,barber2025,ziad2024lcd,kadomoto2025,kasamura2025,itoh2026} and real-time decoders based on machine learning~\cite{senior2025aq2} have been developed to meet this requirement.

For quantum low-density parity-check (qLDPC) codes on superconducting qubits, in contrast, this problem has not yet been sufficiently addressed. Bivariate bicycle (BB) codes, in particular the \BB{144,12,12} ``gross'' code, encode twelve logical qubits in about a tenth of the physical qubits of comparable surface codes~\cite{bravyi2024} and are central to current fault-tolerant roadmaps~\cite{yoder2025tour}. Matching does not apply to them.

\begin{figure}[t]
\centering
\includegraphics{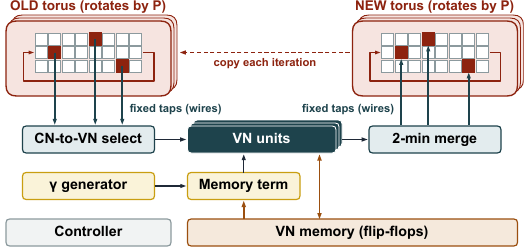}
\caption{RingBP datapath. Each CN orbit keeps its states in an OLD and a NEW register torus that rotate by $P$ slots per step, so the VN units read and write them through fixed taps, which are plain wires.}
\label{fig:arch}
\end{figure}

Among the decoders proposed for BB codes, Relay-BP~\cite{muller2025relay} reaches the accuracy of belief propagation (BP) with ordered-statistics post-processing (BP-OSD)~\cite{panteleev2021,roffe2020} using only local, iterative message passing. It works on a circuit-level decoding matrix whose Tanner graph has tens of thousands of edges. Maurer et al.\ showed that it decodes the gross code in real time on an FPGA~\cite{maurer2025fpga}. Their design is fully parallel. It dedicates a processing element to every variable node (VN) and check node (CN) and finishes one BP iteration in two 12\,ns cycles, at the price of 2.1\,M LUTs (52\% of an AMD XCVU19P) and 58\,W for a single logical block. Because this cost grows with every edge of the graph, and the number of edges is proportional to the code size times the number of syndrome rounds, the fully parallel approach does not extend to larger codes, multi-block systems, or cryogenic power budgets. The fully parallel design is also far faster than a real-time decoder needs to be, since a window of the gross code is decoded in a fraction of a microsecond while a new window arrives only every 8\,\textmu s.

Partially parallel decoding, the standard remedy for classical LDPC codes, does not carry over directly. For a quasi-cyclic code, a few processing elements are time-multiplexed over the circulant blocks, and barrel shifters realize the lift~\cite{mansour2006}. A qubit decoder faces three obstacles. First, the decoder works not on the parity-check matrix of the code, which is quasi-cyclic by construction, but on the circuit-level decoding matrix. This matrix has one column per fault mechanism of the syndrome-extraction circuit, about 60$\times$ more than the code has qubits, and its lift structure must be derived from the circuit. Second, the symmetry of a BB code is a two-dimensional, non-cyclic group, for which a one-dimensional shifter is not sufficient. Third, Relay-BP adds a memory term per node and a random memory strength per node and leg, and this state is expensive to time-multiplex.

This paper presents RingBP, a partially parallel Relay-BP decoder architecture that removes these obstacles with one observation, namely that the circuit-level decoding matrix of a BB memory experiment is a free lift of a small base graph by the automorphism group of the code. For the gross code, the 936$\times$8{,}784 matrix with 30{,}672 edges is a $\mathbb{Z}_6\times\mathbb{Z}_{12}$ lift of a 13$\times$122 base graph with 426 edges, and the base graph grows only linearly with the number of rounds. RingBP turns this lift into wiring. As Fig.~\ref{fig:arch} shows, it builds the base graph in space and sweeps the group in time, while the compressed state of each CN orbit sits in a register torus that rotates in step with the sweep. Every base edge therefore finds its check node at a fixed tap, and no shifter, address table, or crossbar is needed. The folding factor $P$, the number of group elements processed per cycle, trades area for latency. This paper makes the following contributions.

\begin{itemize}
  \item An automorphism-folded architecture, with the lift derived automatically from a Stim circuit, that executes the automorphism group of the decoding graph in time over rotating CN-state tori, so that $P$ scales the datapath without any permutation network.
  \item A hardware-friendly Relay-BP in which one stored sign bit per edge, memory strengths from a stateless jump-ahead RNG, and a bubble-free speculative schedule cut the state and the cycles per iteration.
  \item The first ASIC evaluation of Relay-BP, with 22\,nm and 7\,nm synthesis and activity-annotated power against a full-size fully parallel ASIC and the FPGA, a streaming analysis of deadlines and buffering, and logical error rates on three BB codes.
\end{itemize}

\section{Background}
\label{sec:bg}
\subsection{BB Codes and Decoding Matrices}
A BB decoder works on a circuit-level decoding matrix, which is derived from the code and its syndrome-extraction circuit and is much larger than the code. A BB code is defined by two polynomials $A,B$ over the group algebra of $\gro=\mathbb{Z}_\ell\times\mathbb{Z}_m$. Its check matrices are $H_X=[A\,|\,B]$ and $H_Z=[B^{\!\top}|\,A^{\!\top}]$, so $\gro$ acts on qubits and checks by translation~\cite{bravyi2024}. The gross code uses $\ell=6$ and $m=12$, \BB{72,12,6} uses $6\times6$, and \BB{288,12,18} uses $12\times12$. A memory experiment repeats the syndrome-extraction circuit for $d$ rounds. Under circuit-level noise, each elementary fault flips a set of detectors (comparisons of consecutive syndrome bits). Merging faults with identical signatures yields the circuit-level decoding matrix $H$, a binary matrix of detectors by fault mechanisms, with a prior $p_j$ per column~\cite{gidney2021stim}. Decoding means finding a likely fault set $\hat e$ with $H\hat e=\sigma$ for the observed syndrome $\sigma$. For the gross code with 12 rounds, $H$ has 936 detectors, 8{,}784 fault columns, and 30{,}672 nonzeros.

\subsection{Relay-BP}
Relay-BP~\cite{muller2025relay} extends BP with a memory in every variable node and repeats it in many short runs with different memory strengths. BP passes messages along the edges of the Tanner graph of $H$, which has a VN per fault column and a CN per detector. In each iteration, every VN sends its current belief, a log-likelihood ratio (LLR), to its CNs, and every CN replies to each VN with an estimate computed from its other VNs. In the min-sum variant, the magnitude of this reply is the smallest incoming magnitude. Each VN then adds the replies to its prior LLR $\lambda_j$ to obtain a marginal $M_j$, whose sign gives the estimate $\hat e_j$. Relay-BP replaces the prior of VN $j$ at iteration $t$ by $(1-\gamma_j)\lambda_j+\gamma_j M_j(t-1)$, a mix of the prior and the previous marginal that is weighted by the memory strength $\gamma_j$. Negative and disordered $\gamma_j$ break the trapping sets that stall plain BP. One run with one set of strengths is called a leg. Leg~0 uses a uniform $\gamma_0$, and each later leg draws fresh $\gamma_j$ from an interval and is warm-started from the marginals of the previous leg. Whenever a leg converges ($H\hat e=\sigma$), its solution is kept if its weight $\sum_j\hat e_j\lambda_j$ is the lowest so far, and decoding stops after $S$ solutions or $R$ legs. We use the published gross-code parameters, namely $\gamma_0=0.125$, $\gamma\in(-0.24,0.66)$, 80 iterations in leg~0 and 60 in later legs, $R=60$, and $S=3$.

The FPGA implementation~\cite{maurer2025fpga} fixes the arithmetic, which we adopt so that our comparison isolates the architecture. It uses the int4.2.8 format, that is, sign-magnitude messages with 4-bit magnitudes and priors scaled by 2. The multiplication by the memory strength is a shift-add approximation, and CN outputs are scaled by $\alpha_t=1-2^{-t}$. Each CN is represented by its minimum, second minimum, argmin index, and parity, and each VN keeps a xorshift16 state for drawing $\gamma_j$. In our designs $1-\gamma$ is quantized to 3 fractional bits, and later-leg strengths are drawn from 8 levels.

\subsection{Related Work}
Existing decoder hardware for qLDPC codes is either fully parallel or serial, and neither kind uses the symmetry of the code. Valls et al.\ implemented min-sum and OSD-0 decoders on FPGAs~\cite{valls2021}. Báscones et al.\ explored fully parallel BP+OSD on FPGA and in 45\,nm synthesis and concluded that it becomes impractical beyond distance 12~\cite{bascones2025}. For Relay-BP, the fully parallel FPGA design~\cite{maurer2025fpga} has been compared with OSD and clustering decoders~\cite{maurya2025fpga}, and a min-sum variant is reported in~\cite{wei2026relay}. Closest to our goal, a serially scheduled FPGA decoder for the GARI method reuses hardware on the gross code~\cite{maan2026gari,bascones2026gari}, but its schedule is not derived from the symmetry.

Partially parallel and layered decoders for classical quasi-cyclic LDPC codes~\cite{mansour2006,hocevar2004} assume a designed cyclic lift and store one message per edge. Relative shifting keeps the messages in their shifted position between layers and halves the shifting network~\cite{jang2024}, but a programmable shifter remains. Pipelined layered decoders also stall on data hazards between layers, which a mix of the layered and flooding schedules removes~\cite{petrovic2020}. Layered schedules have also been studied for qLDPC codes~\cite{ducrest2023layered}. RingBP instead handles a two-dimensional lift derived from the circuit, stores no edge message, and needs no shifter.

Automorphisms have so far been used in decoding to improve accuracy, not to reduce hardware. Ensembles of permuted decoders exist for classical~\cite{geiselhart2021} and qLDPC codes~\cite{koutsioumpas2025autdec}. To our knowledge, RingBP is the first Relay-BP decoder architecture for ASICs and the first qubit decoder whose hardware partition is derived from the automorphism group of its decoding graph.

\section{Lift Structure of the Decoding Matrix}
\label{sec:lift}
The circuit-level decoding matrix keeps the translation symmetry of the code, and this symmetry can be found automatically. Let $\mathcal{T}(H)$ be the Tanner graph of $H$, with a node per detector, a node per fault, and an edge per nonzero. We color the fault nodes by their rounded prior and compute the automorphism group of the colored graph with a canonical-labeling tool~\cite{junttila2007bliss}. We apply this procedure to the published Stim circuits~\cite{muller2025relay} for memory experiments with $d$ rounds, the $Z$ basis, and uniform circuit noise. For all three experiments in Table~\ref{tab:codes}, the group is $\gro\cong\mathbb{Z}_A\times\mathbb{Z}_B$ with $(A,B)=(\ell,m)$, generated by two commuting permutations. It acts freely on detectors and faults, that is, every orbit has exactly $|\gro|$ members. The computation takes 4\,s for the gross code and 23\,s for \BB{288,12,18}, including the construction of $H$.

A free action makes $H$ the lift of a small base graph. Choosing one representative per orbit gives every detector a coordinate $(r,g)$, a row orbit $r$ and a group element $g\in\gro$, and every fault a coordinate $(c,g)$. Because the action preserves edges, the edges collapse into base edges $(r,c,s)$, where fault $(c,g)$ is connected to detector $(r,g+s)$ for every $g\in\gro$. $H$ is therefore described by $n_R$ row orbits, $n_C$ column orbits, and $n_E$ base edges, with $n_E\cdot|\gro|=\mathrm{nnz}(H)$. Priors are constant on column orbits, so the LLR of a fault depends only on $c$.

\begin{table}[t]
\centering
\caption{Lift structure of the circuit-level decoding matrices.}
\label{tab:codes}
\footnotesize\setlength{\tabcolsep}{2.5pt}
\resizebox{\columnwidth}{!}{\begin{tabular}{@{}lrrrcrr@{}}
\toprule
Code & rounds & $H$ ($M\times N$) & nnz & $\gro$ & base $n_R\times n_C$ & $n_E$\\
\midrule
\BB{72,12,6}    & 6  & $252\times2{,}232$    & 7{,}776  & $\mathbb{Z}_6\times\mathbb{Z}_6$   & $7\times62$   & 216\\
\BB{144,12,12}  & 12 & $936\times8{,}784$    & 30{,}672 & $\mathbb{Z}_6\times\mathbb{Z}_{12}$ & $13\times122$ & 426\\
\BB{288,12,18}  & 18 & $2{,}736\times26{,}208$ & 91{,}584 & $\mathbb{Z}_{12}\times\mathbb{Z}_{12}$ & $19\times182$ & 636\\
\bottomrule
\end{tabular}}
\end{table}

The lift gives hardware a base graph that is small and has the same node degrees for every code. First, the base graph grows only with the number of rounds ($n_E=216$, 426, and 636 for $d=6$, 12, and 18), whereas $H$ grows with the rounds times $|\gro|$, from 7.8k to 92k edges. Second, the maximum row and column degrees are the same for the three codes (35 and 6), so the node datapaths do not depend on the code.

\section{RingBP Architecture}
\label{sec:arch}
\subsection{Overview}
A partially parallel decoder must choose which dimension of the lift to build in space, and the classical choice of the group is costly for this lift. Classical quasi-cyclic decoders use $|\gro|$ lanes, one per group element, that iterate over the $n_C$ column orbits. Applied to $H$, every cycle would touch different base edges with different shifts, so each lane would need a programmable two-dimensional shifter for $\mathbb{Z}_A\times\mathbb{Z}_B$ and a selector for the row state. In our analysis these fixed costs dominate the area.

RingBP transposes this partition by building the base graph in space and sweeping the group in time (Fig.~\ref{fig:arch}). The base graph is instantiated once, as $n_C$ VN units and $n_E$ edge taps. In step $t$ the hardware processes the $P$ group elements $g(t)+o$, $o\in O$, where $O$ is a fixed set of $P$ offsets, so one iteration takes $|\gro|/P$ steps. The folding factor $P$ thus sets the trade-off, because VN units and taps are replicated $P$ times whereas the CN state is shared by all offsets. In every step, fixed taps read the OLD tori, the CN-to-VN select forms the messages, the VN units add them to the memory term, and the 2-min merge writes the results into the NEW tori, while a controller sequences the steps, iterations, and legs.

\begin{figure}[t]
\centering
\includegraphics[width=\linewidth]{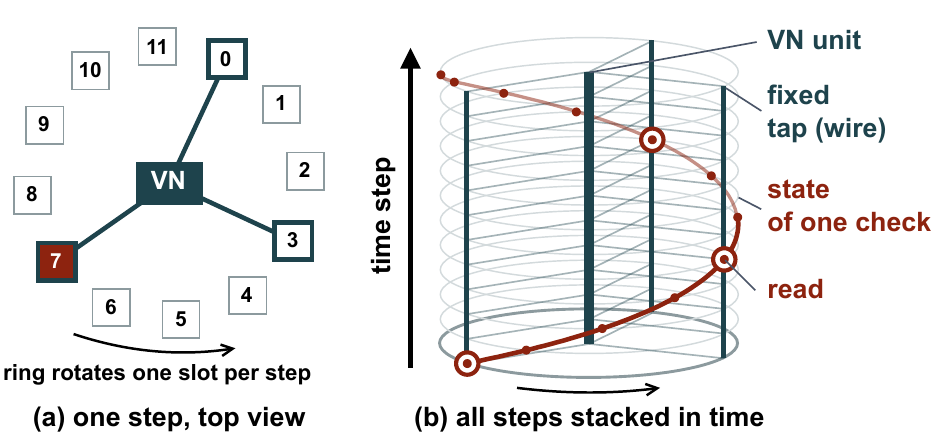}
\caption{Folding of a toy $\mathbb{Z}_{12}$ lift with three base edges. (a) In one step, the VN unit reads three fixed slots of the ring of check states, and the ring then rotates by one slot. (b) Stacked in time, the taps and the VN unit are vertical lines, while the state of a check is a helix that is read where it meets a tap.}
\label{fig:spacetime}
\end{figure}

\subsection{Rotating CN-State Tori}
RingBP moves the storage instead of the address, which turns the lift permutation into fixed wires (Fig.~\ref{fig:spacetime}). With the group in time, base edge $(r,c,s)$ processed for group element $g$ must reach detector $(r,g+s)$, whose identity changes every step. The compressed states of the $|\gro|$ checks of row orbit $r$ (16 bits per check for the minimum, second minimum, argmin slot, and parity) are therefore held in an $A\times B$ register torus. The torus rotates by $P$ along $\mathbb{Z}_B$ every step and by one along $\mathbb{Z}_A$ on wrap-around, which costs one 2:1 multiplexer per bit because $\gro$ is not cyclic. Torus position $q$ then always holds check $(r,g(t)+q)$, so base edge $(r,c,s)$ at offset $o$ always reads and writes position $s+o$.

Two tori per row orbit implement flooding. OLD holds the CN states of iteration $i-1$, from which the CN-to-VN select computes the message of each read tap (exclusive minimum, sign, and $\alpha_t$ scaling). NEW accumulates iteration $i$. Each entry receives the VN-to-CN messages of every base edge that targets it in the current step, merged by a 2-min merge tree with at most 15 inputs in our designs. At the end of an iteration NEW becomes OLD.

\subsection{Variable-Node Memory}
VN state is read and written as one word per step, and no edge message is stored. Each column orbit has one VN unit per offset, with its true degree (2--6) rather than a worst-case degree. Because VN state is accessed in sweep order only, the state of all column orbits for one step forms a single word. This word holds the marginal $M$ (5 bits), $\hat e$, and the best-solution bit of every column orbit, plus one sign bit per base edge. This sign of the VN's previous VN-to-CN message is the only per-edge information needed to remove the VN's own contribution, because the magnitude comes from the OLD torus. With $P$ offsets the word is $P$ times wider and the memory is $|\gro|/P$ words deep (2{,}560\,bits$\times$36 for the gross code at $P=2$).

This organization leaves little state. For the gross code, the decoder holds 92\,kbit of VN memory and 30\,kbit of torus registers, and the synthesized $P=1$ decoder has 128\,k flip-flops against 617\,k for the fully parallel design. %

\subsection{Stateless Memory Strengths}
\label{sec:relay}
Per-node random number generators would undo this saving, so the $\gamma$ generator of RingBP computes the memory strengths without any RNG state. The reference design gives every VN a 16-bit xorshift state to draw $\gamma_j$ for each leg. However, xorshift16 is linear over GF(2). The state after $n$ legs is $x_n=T^n x_0$, and only its top $k=3$ bits are needed to select one of $2^k$ strengths. The controller therefore keeps the three corresponding rows $u_i$ of $T^n$, updates them once per leg ($u_i\leftarrow u_iT$), and broadcasts them. Each VN unit computes $\mathrm{parity}(u_i\wedge x_0)$ with three 16-input AND-XOR trees.

The seed $x_0$ is derived from the lift coordinates, which changes the random draws but not their distribution. It is a hash of the lifted index $l=|\gro|c+g$ of the node, a per-column constant plus a per-step value from a lane ROM. Over 3{,}000 shots, the logical failures with original and lifted seeding do not differ significantly (McNemar $p=0.62$), and the difference is smaller than the variation between RNG seeds.

\subsection{Bubble-Free Schedule}
\label{sec:timing}
RingBP removes all but one of the idle cycles that flooding normally spends between iterations. These cycles serve the convergence test ($H\hat e=\sigma$, an OR over the parities in NEW) and the OLD$\leftarrow$NEW hand-over. (i)~The hand-over happens in the last step, when the tori return to their origin. (ii)~The convergence test is speculative, that is, the next iteration starts at once and the test runs in its step~0 on OLD. If the leg must end, step~0 is aborted by suppressing the memory write enable, and the tori and strengths are re-read in the leg-start state, so an abort costs one cycle per leg rather than one per iteration. (iii)~With a two-stage pipeline (VN unit, merge), one drain cycle remains, because consecutive steps almost always share rows (a row orbit has about 30 base edges whose shifts cover nearly all of $\gro$). A decode thus takes $(|\gro|/P+1)(I+1)+3L$ cycles for $I$ iterations and $L$ legs, which matches RTL simulation exactly.

Two further changes keep the critical path inside the BP datapath. First, the scaling, first-iteration, and initialization controls, which are constant during a sweep, are registered per column instead of being broadcast combinationally. Second, the memory term $\mathrm{sat}(\beta\lambda+\mathrm{sgn}(M)(|M|-\beta|M|))$ with $\beta=1-\gamma$, which depends only on $M$ and $\beta$, is computed one step ahead from an $M$ that is read two steps ahead. The critical path is then torus entry, read tap, VN sum and saturation, and pipeline register, and it does not depend on $P$.

\section{Evaluation}
\label{sec:eval}
\subsection{Experimental Setup}
The workloads are the published Stim circuits for the memory experiments of the three BB codes ($d$ rounds, uniform circuit noise, $Z$-basis detectors)~\cite{muller2025relay}, decoded with the Relay-BP parameters of \S\ref{sec:bg}. Accuracy is compared on identical shots against the open-source floating-point Relay-BP~\cite{muller2025relay}. Our side is the bit-accurate model of the RTL. The RTL is co-simulated in Verilator against this model and matches it bit for bit. Each decoder runs with two RNG seeds, and the comparison uses an exact McNemar test.

All designs are synthesized with Synopsys Design Compiler in a commercial 22\,nm CMOS process and in the ASAP7 7\,nm FinFET predictive kit~\cite{clark2016asap7}. Power is computed on the synthesized netlist during decoding and includes leakage. The switching activity comes from a Verilator simulation of the RTL on sampled syndromes (4{,}000 cycles). The toggle rate and duty of every register bit are annotated and propagated through the logic and the clock gates. Power is evaluated for RingBP only.

The fully parallel (FP) baseline is our RTL implementation of the FPGA microarchitecture~\cite{maurer2025fpga}, with one processing element per VN and CN, two cycles per iteration, and the same arithmetic and I/O. It is synthesized for the full decoding matrix with the same flow.

Latency is computed from the cycle count of the schedule and compared with a per-window budget. A decode takes $(|\gro|/P+1)(I+1)+3L$ cycles (\S\ref{sec:timing}), which we apply to the iteration and leg counts of the bit-accurate model. We decode windows of $d$ rounds and commit $2d/3$ of them, which is a (12,8) window for the gross code. With a 1\,\textmu s syndrome cycle, one window arrives every $2d/3$\,\textmu s, which we call the budget (8\,\textmu s for the gross code).

\subsection{Decoding Accuracy}
\label{sec:acc}
\begin{figure}[t]
\centering
\includegraphics[scale=0.9]{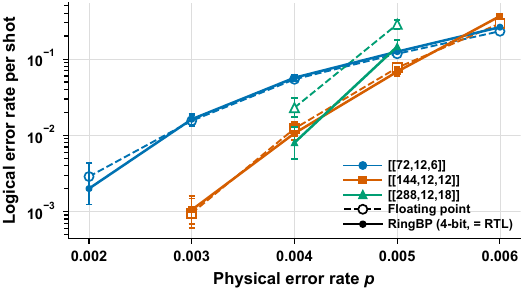}
\caption{Logical error rate per $d$-round memory shot (95\% Wilson intervals; 500--20{,}000 paired shots per point). Solid: RingBP; dashed: floating-point Relay-BP~\cite{muller2025relay}.}
\label{fig:ler}
\end{figure}

The int4 datapath with stateless strengths loses no accuracy in the practically relevant range (Fig.~\ref{fig:ler}). On the gross code the difference to floating-point Relay-BP at $p\le0.005$ is not significant or favors RingBP ($1.06\times10^{-2}$ vs.\ $1.23\times10^{-2}$ at $p=0.004$, and $6.8\times10^{-2}$ vs.\ $7.7\times10^{-2}$ at $p=0.005$ with $p_\mathrm{McNemar}=0.03$). On \BB{72,12,6} the two agree for $p\le0.004$. On \BB{288,12,18} RingBP is better, $8.0\times10^{-3}$ vs.\ $2.3\times10^{-2}$ at $p=0.004$. Within the fixed leg budget, the float decoder needs twice as many iterations on average (651 vs.\ 309) and exhausts the budget more often.

The integer datapath is worse only at high noise, where it converges less often. At $p=0.006$, where the logical error rate exceeds 20\%, it is 1.13--1.25$\times$ worse, and on \BB{72,12,6} it is already 1.07$\times$ worse at $p=0.005$.

\subsection{Area, Speed, and Power}
\label{sec:asp}
Folding does not slow the clock (Table~\ref{tab:gross}). Every RingBP configuration closes timing at 1.05\,ns in 22\,nm and at 1.00\,ns in 7\,nm with no violation, because the critical path stays inside one VN unit and its taps. The FP design is slower per cycle and closes timing at 1.41\,ns in 22\,nm and at 1.20\,ns in 7\,nm.

Folding provides design points that are 4--12$\times$ smaller than FP. RingBP needs 0.36\,mm$^2$ at $P=1$, 8.6\% of FP's 4.24\,mm$^2$, and 1.02\,mm$^2$ at $P=12$ (24\%), while the time per iteration falls from 77\,ns to 7.4\,ns. FP remains the fastest per iteration (2.8\,ns). Its mean latency of 0.08\,\textmu s, however, is 100$\times$ below the 8\,\textmu s budget, and this slack is what folding converts into area. The latency column gives the mean and the 99th percentile of the decoding time of a 12-round window at $p=10^{-3}$, over $10^7$ shots in 22\,nm. A decode averages 12.2 iterations in three legs, and the FP latency includes $5L+5C+2$ control cycles for $C$ converged legs besides its $2I$ iteration cycles. The area--time product of FP is better than that of RingBP for $P\le2$ and equal within 7\% at $P=4$, and RingBP is better at $P=6$ and 12 (0.77$\times$ and 0.63$\times$). In 7\,nm the area is 5.5--6.1$\times$ smaller.

\begin{table}[t]
\centering
\caption{Synthesis results for the gross code.}
\label{tab:gross}
\footnotesize\setlength{\tabcolsep}{2.2pt}
\resizebox{\columnwidth}{!}{\begin{tabular}{@{}lrrrrrrrrrrr@{}}
\toprule
 & & & \multicolumn{6}{c}{22\,nm} & \multicolumn{2}{c}{7\,nm
 } & latency\\
\cmidrule(lr){4-9}\cmidrule(lr){10-11}
 & cyc/ & FF & area & rel. & ns/ & A$\cdot$T & power & nJ/ & area & power & mean / p99\\
 & iter & (k) & mm$^2$ & & iter & & mW & iter & mm$^2$ & mW & \textmu s\\
\midrule
FP        & 2  & 617 & 4.24  & 1     & 2.8 & 12.0 & --- & --- & 0.712 & --- & 0.08 / 0.17\\
$P{=}1$   & 73 & 128 & 0.364 & 0.086 & 76.7 & 27.9 & 174 & 13.3 & 0.066 & 95  & 1.02 / 3.38\\
$P{=}2$   & 37 & 133 & 0.428 & 0.101 & 38.9 & 16.6 & 204 & 7.9  & 0.076 & 113 & 0.52 / 1.72\\
$P{=}4$   & 19 & 143 & 0.556 & 0.131 & 20.0 & 11.1 & 279 & 5.6  & 0.097 & 158 & 0.27 / 0.89\\
$P{=}6$   & 13 & 153 & 0.671 & 0.158 & 13.7 & 9.2  & 326 & 4.4  & 0.115 & 190 & 0.19 / 0.61\\
$P{=}12$  & 7  & 182 & 1.019 & 0.240 & 7.4  & 7.5  & 497 & 3.7  & 0.168 & 282 & 0.11 / 0.33\\
\bottomrule
\end{tabular}}\\[2pt]
\footnotesize A$\cdot$T is the area--time product in mm$^2\cdot$ns.
\end{table}

Power grows more slowly than throughput. During decoding, RingBP draws 174--497\,mW in 22\,nm, of which 41--49\% is the clock network that shifts the tori. The energy per iteration therefore falls from 13.3\,nJ at $P=1$ to 3.7\,nJ at $P=12$. In 7\,nm the power is 95--282\,mW.

Against the FPGA, which runs the same algorithm in the same number format, RingBP is faster for $P\ge4$ and uses two orders of magnitude less energy (Table~\ref{tab:fpga}). One iteration takes 20.0\,ns at $P=4$ and 7.4\,ns at $P=12$ against 24\,ns. The power is 0.28--0.50\,W against 58\,W, and the energy per iteration is 5.6 and 3.7\,nJ against 1.39\,\textmu J, that is, 250--380$\times$ lower. Because the comparison spans technologies and estimation methods (an FPGA tool estimate against synthesis-level ASIC power), the ratio should be read as an order of magnitude.

\begin{table}[t]
\centering
\caption{Comparison with the FPGA decoder for the gross code.}
\label{tab:fpga}
\footnotesize\setlength{\tabcolsep}{5pt}
\begin{tabular}{@{}lccc@{}}
\toprule
 & FPGA~\cite{maurer2025fpga} & RingBP $P=4$ & RingBP $P=12$\\
\midrule
Technology & 16\,nm FPGA & \multicolumn{2}{c}{22\,nm ASIC}\\
Architecture & fully parallel & \multicolumn{2}{c}{automorphism-folded}\\
Number format & int4.2.8 & \multicolumn{2}{c}{int4.2.8}\\
Logic & 2.11\,M LUT & 0.56\,mm$^2$ & 1.02\,mm$^2$\\
Flip-flops & 0.54\,M & 143\,k & 182\,k\\
Clock period & 12\,ns & 1.05\,ns & 1.05\,ns\\
Cycles per iteration & 2 & 19 & 7\\
Time per iteration & 24\,ns & 20.0\,ns & 7.4\,ns\\
Power & 58.0\,W & 0.28\,W & 0.50\,W\\
Energy per iteration & 1.39\,\textmu J & 5.6\,nJ & 3.7\,nJ\\
Latency, mean / p99 & --- & 0.27 / 0.89\,\textmu s & 0.11 / 0.33\,\textmu s\\
\bottomrule
\end{tabular}
\end{table}

\subsection{Streaming Decoding}
\label{sec:stream}
Real-time operation is decided by the tail of the latency, not by its mean. Relay-BP's decoding time has a heavy tail, and the treatment of slow windows determines both the logical error rate and the input buffer. We therefore decode a syndrome stream. Memory experiments of up to 75 rounds, built by repeating the middle rounds of the published circuit, are decoded with (12,8) sliding windows~\cite{skoric2023,gong2024gdg} by the bit-accurate model. Every window matrix is again a $\mathbb{Z}_6\times\mathbb{Z}_{12}$ lift (864$\times$8{,}424, base graph 12$\times$117 with 404 edges), so the same hardware applies. Without a deadline, the per-round logical error rate of the stream is $1.9\times10^{-7}$, $6.3\times10^{-6}$, and $1.1\times10^{-4}$ at $p=1$, 2, and $3\times10^{-3}$, which is 1.05--1.39$\times$ that of single-shot 12-round decoding.

We model a deadline $D$ and the input buffer that it implies. A window is aborted $D$ after its decoding starts, and the lowest-weight solution found so far (or the hard decision of leg~0 if there is none) is committed. Because a wrong commit propagates to later windows, we count failures over whole streams and re-decode the affected streams with the same random numbers. The buffer depth is obtained from a queue with one arrival per 8\,\textmu s and the measured decoding times, for an overflow probability below $10^{-6}$. For the 22\,nm design at 1.05\,ns, each entry of Table~\ref{tab:stream} gives the per-round logical error rate relative to no deadline and the buffer depth in windows, and a dash marks a buffer deeper than 32 windows.

A small $P$ must pay for the tail with accuracy or with buffering. A strict 8\,\textmu s deadline needs no buffer but multiplies the logical error rate by 24, 8.8, and 2.9 for $P=2$, 4, and 12 at $p=10^{-3}$, almost entirely through windows that are aborted before their first solution. $P=2$ recovers to 1.6--1.9$\times$ with $D=64$\,\textmu s and 14--21 buffered windows. $P=4$ reaches parity (1.0--1.1$\times$) at $D=64$\,\textmu s with 14--15 windows for $p\le2\times10^{-3}$. Exempting solution-less windows from the deadline restores the accuracy for any $D$ (within 1.3$\times$) but brings the tail back. $P=2$ then needs more than 30 windows and $P=4$ needs 16--17.

At $P=12$ the deadline never binds. This decoder exhausts the whole budget of 60 legs in 27\,\textmu s, so with $D=32$\,\textmu s no window is ever truncated. The logical error rate is identical to the unconstrained one up to $p=3\times10^{-3}$, with a buffer of 4--8 windows and a load of 0.02--0.12. The 7\,nm results are nearly identical. Folding thus leaves a real choice between $P=4$ at 0.56\,mm$^2$, if a 64\,\textmu s reaction time is acceptable, and $P=12$ at 1.02\,mm$^2$ for 32\,\textmu s with no accuracy loss.

\begin{table}[t]
\centering
\caption{Streaming decoding of the gross code under a deadline $D$.}
\label{tab:stream}
\footnotesize\setlength{\tabcolsep}{4pt}
\begin{tabular}{@{}lrccc@{}}
\toprule
$p$ & $D$ (\textmu s) & $P=2$ & $P=4$ & $P=12$\\
\midrule
$10^{-3}$ & 16 & 8.5$\times$ / 2 & 3.7$\times$ / 2 & 1.3$\times$ / 2\\
          & 32 & 3.7$\times$ / 6 & 1.7$\times$ / 6 & 1.0$\times$ / 4\\
          & 64 & 1.6$\times$ / 14 & 1.0$\times$ / 14 & 1.0$\times$ / 4\\
\midrule
$2\times10^{-3}$ & 16 & 9.5$\times$ / 4 & 4.2$\times$ / 3 & 1.5$\times$ / 2\\
          & 32 & 4.0$\times$ / 10 & 1.9$\times$ / 8 & 1.0$\times$ / 5\\
          & 64 & 1.9$\times$ / 21 & 1.1$\times$ / 15 & 1.0$\times$ / 5\\
\midrule
$3\times10^{-3}$ & 16 & 7.0$\times$ / 7 & 3.7$\times$ / 5 & 1.5$\times$ / 4\\
          & 32 & 3.6$\times$ / 22 & 1.9$\times$ / 14 & 1.0$\times$ / 8\\
          & 64 & 1.9$\times$ / --- & 1.1$\times$ / 31 & 1.0$\times$ / 8\\
\bottomrule
\end{tabular}
\end{table}

\subsection{Ablation Study}
\label{sec:abl}
The area saving comes from state. The transposed partition at $P=1$ holds 128\,k flip-flops, 4.8$\times$ fewer than FP's 617\,k, because no edge message is stored and the CN state is compressed. Each additional offset adds only one slice of the VN word, so the flip-flop count grows by 42\% from $P=1$ to $P=12$, while the area grows 2.8$\times$ with the replicated VN units and merge trees. The stateless RNG is essential here. The per-VN xorshift state of the reference design would add $16\,n_C=1{,}952$ bits to each of the 72 words, 141\,kbit in total, and more than double the state of the $P=1$ decoder.

The schedule and critical-path techniques of \S\ref{sec:timing} cut the time per iteration by about a third (Table~\ref{tab:ablation}). We isolate them on the one-round sub-problem at $P=36$ with a two-stage pipeline, synthesize each design over a sweep of clock targets, and report the smallest achieved period (target minus worst slack) with the area in mm$^2$ at that point. Removing the iteration bubbles saves one of four cycles per iteration. The per-column control registers and the memory-term precompute then shorten the period by 5\% in 22\,nm and 10\% in 7\,nm. The precompute also removes 8\% of the flip-flops, because the per-unit $\beta$ and base registers become unnecessary. Together, the time per iteration falls from 3.9 to 2.7\,ns in 22\,nm and from 3.1 to 2.1\,ns in 7\,nm. On this small sub-problem, FP remains 1.2$\times$ faster per iteration at 1.8--1.9$\times$ the area.

\begin{table}[t]
\centering
\caption{Timing ablation on the one-round sub-problem.}
\label{tab:ablation}
\footnotesize\setlength{\tabcolsep}{2.6pt}
\begin{tabular}{@{}lrrrrrrr@{}}
\toprule
 & & \multicolumn{3}{c}{22\,nm} & \multicolumn{3}{c}{7\,nm}\\
\cmidrule(lr){3-5}\cmidrule(l){6-8}
Design & cyc/it & area & ns & ns/it & area & ns & ns/it\\
\midrule
Fully parallel & 2 & 0.519 & 1.12 & 2.24 & 0.086 & 0.90 & 1.80\\
RingBP with bubbles & 4 & 0.310 & 0.98 & 3.92 & 0.044 & 0.77 & 3.09\\
\ + bubble-free check & 3 & 0.311 & 0.94 & 2.82 & 0.044 & 0.77 & 2.32\\
\ + per-column control & 3 & 0.313 & 0.93 & 2.79 & 0.045 & 0.71 & 2.13\\
\ + memory-term precomp. & 3 & 0.287 & 0.89 & 2.67 & 0.045 & 0.69 & 2.07\\
\bottomrule
\end{tabular}
\end{table}

\subsection{Scaling to Other Codes}
\label{sec:scale}
\begin{figure}[t]
\centering
\includegraphics[scale=0.9]{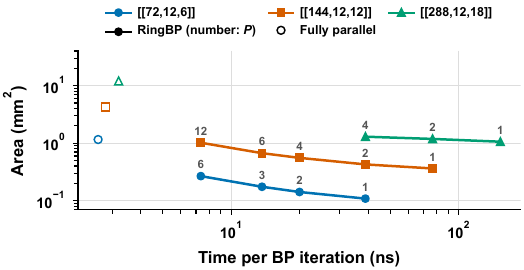}
\caption{Area vs.\ time per iteration for three BB codes (22\,nm synthesis). Filled markers show RingBP with $P$ as the label, and open markers show the fully parallel ASIC.}
\label{fig:scaling}
\end{figure}

The area saving holds over a 12$\times$ range of matrix size but does not grow with it (Fig.~\ref{fig:scaling}). The FP area follows the number of edges of $H$, with 1.16, 4.24, and 12.1\,mm$^2$ in 22\,nm for the three codes. RingBP at $P=1$ occupies 0.109, 0.364, and 1.06\,mm$^2$ at the same 1.05\,ns clock, which is 9.4\%, 8.6\%, and 8.8\% of FP. The ratio stays flat because, with flip-flop memories, the state of RingBP (35\,k to 373\,k flip-flops) grows with the number of columns of $H$, as that of FP does.

A larger code needs a larger $P$ to stay within its budget, at a small area cost. At $p=10^{-3}$, $P=1$ decodes every sampled window of the two smaller codes within the budget (worst case 2.9 and 6.0\,\textmu s against 4 and 8\,\textmu s). For \BB{288,12,18}, 0.6\% of the windows exceed the 12\,\textmu s budget at $P=1$. $P=2$ (1.19\,mm$^2$, 9.8\% of the FP area) has a worst case of 8.9\,\textmu s, and $P=4$ (1.30\,mm$^2$, 10.7\%) of 4.5\,\textmu s.

\section{Conclusion}
This paper presented RingBP, a partially parallel Relay-BP decoder for BB codes. Its key observation is that the symmetry of the code survives into the circuit-level decoding matrix as a free group lift, and rotating CN-state tori turn this lift into fixed wiring. For the gross code, RingBP occupies 0.36--1.02\,mm$^2$ in 22\,nm, 9--24\% of a fully parallel ASIC, at a clock period that does not depend on the folding factor. At $P=12$ it runs an iteration in 7.4\,ns at 0.50\,W and decodes a syndrome stream under a 32\,\textmu s deadline with no loss in logical error rate, and the same architecture covers BB codes up to $[[288,12,18]]$. 

\bibliographystyle{IEEEtran}
\bibliography{refs}

\end{document}